\documentclass[a4paper,10pt]{article}
\usepackage{multicol}
\usepackage{amsmath}
\usepackage{amssymb}
\usepackage{graphicx}
\usepackage{float}
\usepackage{amsmath}
\usepackage{bm}
\usepackage{amsfonts}
\usepackage{balance}
\usepackage{titling}
\usepackage{cite}
\usepackage{color}
\makeatletter
\newcommand\figcaption{\def\@captype{figure}\caption}
\makeatother
\newcommand{\bee}{\begin{equation}}
\newcommand{\ee}{\end{equation}}
\newcommand{\beea}{\begin{eqnarray}}
\newcommand{\eea}{\end{eqnarray}}

 \newlength{\halfpagewidth}
 \divide\halfpagewidth by 2
\newcommand{\leftsep}
{\noindent\raisebox{4mm}[0ex][0ex]{\makebox[\halfpagewidth]{\hrulefill}\hbox{\vrule height 3pt}}                         \vspace*{-2mm}                }
\newcommand{\rightsep}
{\noindent\hspace*{\halfpagewidth}\rlap{\raisebox{-3pt}[0ex][0ex]{\hbox{\vrule height 3pt}}}                         \makebox[\halfpagewidth]{\hrulefill}  }
\thanksfootextra{}{)}
\thanksheadextra{}{)}
\thanksmarkseries{alph}
\begin{document}
\title{Low-energy quantum scattering induced by graphene ripples}
\author{ {Daqing Liu$^{1,3}$\thanks{Corresponding author: liudq@cczu.edu.cn}, Xin Ye$^{1}$, Shuyue Chen$^{1}$, Shengli Zhang$^3$, Ning Ma$^{2,3}$\thanks{Corresponding author: maning@stu.xjtu.edu.cn}} \\
{\small \it $^{1}$ School of Mathematics and Physics, Changzhou University, Changzhou, 213164, China}\\
{\small \it $^2$ Department of Physics, Taiyuan University of Technology, Taiyuan, 030024, China}\\
{ \small \it $^3$ School of Science, Xi'an Jiaotong University, Xi'an, 710049, China }\\
 }
\date{}
\maketitle





 \begin{abstract}
We report a quantum study of the carrier scattering induced by graphene ripples. Crucial differences between the scattering induced by the ripple and ordinary scattering were found. In contrast to the latter, in which the Born approximation is valid for high-energy process, the former is valid for the low-energy process with a quite broad energy range. Furthermore, in polar symmetry ripples, the scattering amplitude exhibits a pseudo-spin structure, an additional factor $\cos\theta/2$, which leads to an absence of backward scattering. We also elucidate that the scattering cross sections are proportional to the energy cubed of the incident carrier.

 \end{abstract}

{\bf keywords:} Graphene ripple, Quantum scattering, Born approximation

\begin{multicols}{2}

\section{Introduction}

Graphene\cite{graphene,graphene1}, which is a single atomic layer of carbon, is a two-dimensional crystal that can be embedded into a three-dimensional space. In the past decade, its isolation has attracted a substantial amount of theoretical and experimental research,  
 most of which stemmed from the peculiar behavior of graphene carriers. Many applications focus on the flat graphene, in which the movement of carriers was represented by the Dirac Hamiltonian with massless fermion in 2+1 dimensional space-time\cite{rmd,wallace,zero1}.

Iorio et al. \cite{lorio1,lorio2,hawking,lorio3,lorio4} studied the Weyl-gauge symmetry of graphene Hamiltonian and its application to gravity research. Their results revealed that if one accepts the flat space-time description of conduct electron, one must also accept a curved space-time description because through a Weyl redefinition of the fields, actions are the same.

However, the Dirac description of graphene is only valid for low energies and small momentum around the Dirac points $ka_l <<1$, where $a_l\sim 10^{-10}m$ is the nearest carbon-to-carbon distance. In other words, the Dirac description of graphene is valid for carrier energies satisfying $E=\hbar v_F k \ll E_l=\hbar v_F/a_l\simeq 3 \text{eV}$. Then, in the following, we restrict all the discussions in the constraints  $|E| \ll 3 \text{eV}$ around the Dirac points.

It is well known that both experimental and theoretical studies have revealed that graphene is always corrugated and covered by ripples, which can be either intrinsic\cite{intr1,intr2} or induced by roughness of substrate\cite{subs1,subs2}. Although graphene ripples complicate the system, they also extend their application because additional adjustable parameters are associated with the ripples\cite{graphene1,ripple1}.

For instance, incident carriers are scattered or deflected by the nonzero curvature induced by graphene ripples. However, in such process the quantum behaviors of the carriers should be considered. In fact, many questions remain, such as the role of the pseudospin in the scattering process.

Katsnelson and Geim \cite{katsnelson} have studied the influence on the electronic quality by the corrugation scattering effect using a fractal dimension theory.  They showed that when the corrugation graphene can be depicted by a fractal dimensional surface, {\it i.e.} the height-correlation function behaviors as $<[(h[\mathbf{r}]-h(0))^2>\propto r^{2H}$, the effective potential is a long-range one, such as Eq. (15) in Ref. \cite{katsnelson}. However, when graphene ripples can not be depicted by fractal dimension theory, the detail study on the quantum scattering induced by, for instance, isolated ripple, is needed.

In this manuscript we report a quantum study on the carrier scattering induced by graphene ripples.
Surprisingly, our study showed a crucial difference between the scattering induced by the graphene ripple and ordinary scattering. In the latter, the Born approximation was valid in the high-energy process, whereas in the scattering by typical graphene ripples, the Born approximation showed an opposite validity
but with a quite broad energy range provided the Dirac description of graphene is valid. We furthermore showed that, in a polar symmetry ripple, the scattering process emerged an explicit pseudospin structure, in which an additional factor $\cos\theta/2$ was exhibited in the expression of scattering amplitude, which leads to the absence of backward scattering. Compared to ordinary quantum scattering, the scattering cross sections were proportional to the energy cubed of the incident carrier. Furthermore, the Hamiltonian also showed that if the graphene was bent in only one direction, its properties did not change if there is no phase transition because  suitable parameters could be rearranged.

\section{Hamiltonian in the ripple graphene}
In this study, we assume that the graphene is not consistently flat, that is, the curvature of graphene is not zero. To depict the curvature, we first formulate graphene metric.

A curve graphene can be considered a two-dimensional surface embedded in a three-dimensional space. The surface is defined by the function $z(\mathbf{r})$, which is the height with respect to the $z=0$ plane and $\mathbf{r}=(x,y)$ is the coordination in $z=0$ plane, as shown in Eq. (\ref{dz}):
\bee \label{dz}
dz^2=z_x^2dx^2+z_y^2dy^2+2z_xz_ydxdy,
\ee
where $z_x=\frac{\partial z}{\partial x}$, $z_y=\frac{\partial z}{\partial y},$ $z_{xy}=\frac{\partial^2 z}{\partial x \partial y}$ etc.

The space-time metric is (in this study we set $\hbar=v_F\equiv 1$)
\bee \label{metric}
g_{\mu\nu}=\left(
\begin{array}{ccc}
1 & 0 & 0 \\
0 & -\left(z_x(x,y)^2+1\right) & -z_x(x,y) z_y(x,y) \\
0 & -z_x(x,y) z_y(x,y) & -\left(z_y(x,y)^2+1\right) \\
\end{array}
\right),
\ee
where the zeroth component on 2+1 dimensional space-time, $t$, does not mix with space components. Thus, in the following, the zeroth component is ignored if there is no confusion.

The fielbein fields $e^a_{\cdot\mu}(x,y)$, which satisfy $e^a_\mu e^b_\nu \eta_{ab} =g_{\mu\nu}$ with the constant metric $\eta_{a,b}=diag\{-1,-1\}$ is chosen as $e^1_{\cdot 2}=e^2_{\cdot 1}$,
\bee
e^a_{\cdot\mu}=
\left(
\begin{array}{cc}
 \frac{z_x^2\sqrt{z_x^2+z_y^2+1} +z_y^2}{z_x^2+z_y^2} & \frac{z_x z_y \left(\sqrt{z_x^2+z_y^2+1}-1\right)}{z_x^2+z_y^2} \\
\frac{z_x z_y \left(\sqrt{z_x^2+z_y^2+1}-1\right)}{z_x^2+z_y^2} &  \frac{z_y^2\sqrt{z_x^2+z_y^2+1} +z_x^2}{z_x^2+z_y^2}
\end{array}
\right).
\ee
 This choice guarantees that when graphene is flat, fielbein matrices are consistently unitary transformations.

Because  the affine connection are defined by
$\Gamma^{\mu}_{\nu\lambda}= \frac{1}{2}g^{\rho\mu}
(g_{\rho\nu,\lambda}+g_{\rho\lambda,\nu}-g_{\nu\lambda,\rho}), $
 the affine connections are read as
\beea \label{affine}
\Gamma_1 &=& \Gamma^\mu_{1\nu}=
\left(
\begin{array}{cc}
 \frac{z_x z_{xx}}{1+z_x^2+z_y^2} & \frac{z_x z_{xy}}{1+z_x^2+z_y^2} \\
 \frac{z_y z_{xx}}{1+z_x^2+z_y^2} & \frac{z_y z_{xy}}{1+z_x^2+z_y^2} \\
\end{array}
\right), \notag \\
\Gamma_2 &=& \Gamma^\mu_{2\nu}=
\left(
\begin{array}{cc}
 \frac{z_x z_{xy}}{1+z_x^2+z_y^2} & \frac{z_x z_{yy}}{1+z_x^2+z_y^2} \\
 \frac{z_y z_{xy}}{1+z_x^2+z_y^2} & \frac{z_y z_{yy}}{1+z_x^2+z_y^2} \\
\end{array}
\right).
\eea

In flat space, the gamma matrices are
\beea
\gamma^{\dot{1}} &=&
\left(
\begin{array}{cc}
0  & 1 \\
-1 & 0
\end{array}
\right),\, ~~
\gamma^{\dot{2}} =
\left(
\begin{array}{cc}
0  & -i \\
-i & 0
\end{array}
\right), \notag \\
\gamma^0 &=&
\beta=
\left(
\begin{array}{cc}
1  & 0 \\
0  & -1
\end{array}
\right).
\eea
 Whereas in ripple graphene they are defined as $\gamma^\mu=e^\mu_a \gamma^a $,
\beea
\gamma^1 &=&
\left(
\begin{array}{cc}
 0 & \frac{\frac{z_x}{\sqrt{1+z_x^2+z_y^2}}+i z_y}{z_x+i z_y} \\
 \frac{i z_y - \frac{z_x}{\sqrt{1+z_x^2+z_y^2}}}{z_x-i z_y} & 0 \\
\end{array}
\right),
 \notag \\
\gamma^2 &=&
\left(
\begin{array}{cc}
 0 & \frac{\frac{z_y}{\sqrt{1+z_x^2+z_y^2}}- i z_x}{z_x+i z_y} \\
 \frac{-i z_x - \frac{z_y}{\sqrt{1+z_x^2+z_y^2}}}{z_x-i z_y} & 0 \\
\end{array}
\right).
\eea

When the spin connection coefficient is defined as $\omega_\lambda^{ab}=e^{a\sigma}(e^b_{\cdot\sigma,\mu}-\Gamma^\lambda_{\mu\sigma}e^b_{\cdot\lambda})$, and the spin connection is defined as $\Omega_\mu=\frac{1}{8}\omega^{ab}_\mu [\gamma_a,\gamma_b]$, we find
\beea
\Omega_1 &=&
\Omega
\left(
\begin{array}{cc}
\frac{z_x z_{xy}-z_y z_{xx}}{z_x^2+z_y^2} & 0 \\
0 & \frac{z_y z_{xx}-z_x z_{xy}}{z_x^2+z_y^2} \\
\end{array}
\right), \notag \\
\Omega_2 &=&
\Omega
\left(
\begin{array}{cc}
\frac{z_x z_{yy}-z_y z_{xy}}{z_x^2+z_y^2} & 0 \\
0 & \frac{z_y z_{xy}-z_x z_{yy}}{z_x^2+z_y^2} \\
\end{array}
\right),
\eea
with $\Omega= \frac{i (\sqrt{1+z_x^2+z_y^2}-1) }{2  \sqrt{1+z_x^2+z_y^2}}$.

\end{multicols}
\leftsep

The Hamiltonian is a very complex $2\times 2 $ matrix, as shown in Eq. (\ref{hamt}),

\beea \label{hamt}
H_{11} &=& H_{22}=0, \notag \\
H_{12} &=&
 -\frac{i}{z_x+i z_y} \{ (-i z_x+\frac{z_y}{\sqrt{1+z_x^2+z_y^2}}) (\partial_y-\frac{i (\sqrt{1+z_x^2+z_y^2}-1) (z_x z_{yy}-z_{xy} z_y)}{2 (z_x^2+z_y^2) \sqrt{1+z_x^2+z_y^2}}) + \notag \\ &&
(i z_y+\frac{z_x}{\sqrt{1+z_x^2+z_y^2}}) (\partial_x-\frac{i (z_x z_{xy}-z_{xx} z_y) (\sqrt{1+z_x^2+z_y^2}-1)}{2 (z_x^2+z_y^2) \sqrt{1+z_x^2+z_y^2}})\}, \notag \\
H_{21} &=&
 -\frac{i}{z_x-i z_y} \{ ((i z_x+\frac{z_y}{\sqrt{1+z_x^2+z_y^2}}) (\partial_y+\frac{i (\sqrt{1+z_x^2+z_y^2}-1) (z_x z_{yy}-z_{xy} z_y)}{2 (z_x^2+z_y^2) \sqrt{1+z_x^2+z_y^2}}) -\notag \\ && (i z_y-\frac{z_x}{\sqrt{1+z_x^2+z_y^2}}) (\partial_x+\frac{i (z_x z_{xy}-z_{xx} z_y) (\sqrt{1+z_x^2+z_y^2}-1)}{2 (z_x^2+z_y^2) \sqrt{1+z_x^2+z_y^2}})) .
\eea
However, the complexity of the Hamiltonian should not be intimidating. If graphene is only bent in x-direction, that is, $z_y=z_{xy}=z_{yy}=0$, the Hamiltonian becomes
$$
H^\prime = \left(
\begin{array}{cc}
 0 & -\partial_y-\frac{i \partial_x}{\sqrt{1+z_x^2}} \\
 \partial_y-\frac{i \partial_x}{\sqrt{1+z_x^2}} & 0 \\
\end{array}
\right).
$$
In the above expression,  if we make a substitution $x \to x^\prime$ and $y \to y^\prime$, where $y^\prime =y$ and $x^\prime=\int^x_0 \sqrt{1+z_x^2} dx$, the Hamiltonian is the same as the one in flat graphene. The physics effect is a Fermi velocity renormalization in x-direction,
$v_F \to \frac{v_F}{\sqrt{1+z_x^2}}$.

Such bend leads to an effective potential. To illustrate this point we consider a bent graphene depicted by
$z(x)=\left\{\begin{array}{ll} 0, & \mathbf{I}:x \le 0 \\ c x, & \mathbf{II}:0<x<d \\ cd, & \mathbf{III}:x \ge d\end{array}\right. $ and normal incident conduction electrons (with energy $k>0$) in the left side. In this case $z_x=c$ at region $\mathbf{II}$ and $z_x=0$ at regions $\mathbf{I}$ and $\mathbf{III}$. Wave functions in three regions are
\[
\psi_I=\frac{1}{\sqrt{2}}\left(\begin{array}{l} 1 \\1 \end{array}\right) e^{ikx}+r
\frac{1}{\sqrt{2}}\left(\begin{array}{l} 1 \\ -1 \end{array}\right) e^{-ikx},
\]
\[
\psi_I=a \frac{1}{\sqrt{2}}\left(\begin{array}{l} 1 \\1 \end{array}\right) e^{i\sqrt{1+c^2}kx}+r
\frac{1}{\sqrt{2}}\left(\begin{array}{l} 1 \\ -1 \end{array}\right) e^{-i\sqrt{1+c^2}kx}
\]
and
\[\psi_{III}=t \frac{1}{\sqrt{2}}\left(\begin{array}{l} 1 \\1 \end{array}\right) e^{ikx},
\]
respectively. Comparing above results with Eqs. (25)-(27) in Ref. \cite{rmd}, we find that the bend leads to an effective square well potential,
\[
V(x)=\left\{\begin{array}{ll} -(\sqrt{1+c^2}-1)k, & 0<x<d, \\ 0, & \text{otherwise}. \end{array}\right.
\]
One has a similar result obviously at the case of oblique incidence.

For simplification, we assume in the following that the ripple is polar symmetrical, that is, in the polar coordinates, $z_\theta=0$, where $\theta$ is the polar angle with respect to the x-axis.

In this situation Hamiltonian turns into
\bee \label{hm1}
H=
\left(
\begin{array}{cc}
 0 & e^{-i\theta} [ -\frac{i}{\sqrt{1+z_r^2}}\partial_r -\frac{\partial_\theta}{r}+\frac{i}{2r}(1-\frac{1}{\sqrt{1+z_r^2}})  ]   \\
e^{i\theta} [ -\frac{i}{\sqrt{1+z_r^2}}\partial_r +\frac{\partial_\theta}{r}+\frac{i}{2r}(1-\frac{1}{\sqrt{1+z_r^2}})  ] & 0 \\
\end{array}
\right)
\ee

\rightsep
\begin{multicols}{2}

In the following, we assume that the ripple is not only small but also local, that is, $z_r \ll 1$ at $r<a$ and $z_r$ drops off rapidly at $r \ge a$. The Hamiltonian can then be decomposed as,
\bee
H=H_0+V,
\ee
where
\bee
H_0=
\left(
\begin{array}{cc}
0  &  e^{-i\theta}(-i\partial_r-\frac{\partial_\theta}{r})   \\
e^{i\theta}(-i\partial_r+\frac{\partial_\theta}{r})  & 0
\end{array}
\right)
\ee
is the Hamiltonian in flat graphene
  and an effective potential induced by graphene ripples
\bee
V=\frac{i z_r^2}{2}
\left(
\begin{array}{cc}
0 & e^{-i\theta}(\frac{1}{2r}+\partial_r) \\
e^{i\theta}(\frac{1}{2r}+\partial_r) & 0
\end{array}
\right).
\ee

There are $\partial$-terms in the effective potential, which will cause scattering effect proportional to $ka$,
 which has two results: First, in a large momentum the perturbational treatment is invalid; second, the unitarity of the result should be broken, that is, there will be no the optical theorem in the scattering process.


Notice that if we make a unitary transformation (FWT transformation) $\mathcal{U}$, the Hamiltonian $H_0$ will turn into $H_0^\prime=\mathcal{U} H_0 \mathcal{U}^\dag$. Particularly, by letting
$\mathcal{U}=e^{\frac{i}{2}\theta\sigma_3}=\left( \begin{array}{cc} e^{\frac{i\theta}{2}} & 0 \\ 0 & e^{-\frac{i\theta}{2}} \end{array}\right)
$, where $\sigma_3$ is the generator of the rotation with respect to z-axis, we have
\[H^\prime_0=-i \left( \begin{array}{cc} 0 & \partial_r+i\frac{\partial_\theta}{r}+\frac{1}{2r} \\ \partial_r-i\frac{\partial_\theta}{r}+\frac{1}{2r} & 0 \end{array}\right),
\]
which is the same as that in Refs. \cite{prb07,report} (for instance, Eq. (9) in Ref. \cite{prb07}) up to a factor $-i$.
Therefore, although our Hamiltonian is not the same as the one given in Refs.\cite{prb07,report}, they are only up to a FWT transformation. In the present study, the base of pseudo-spinor is a constant vector
$
\left(
\begin{array}{c}
\psi_A \\ \psi_B
\end{array}
\right),
$
whereas in Refs.\cite{prb07,report} the base of pseudo-spinor is a position-dependent vector.

From the expression of $H_0$, the wave function, which represents a particle with energy $k>0$ moving along $\hat{k}=(\cos\theta_0,\sin\theta_0)$ is
\bee \label{wave1}
\psi(r,\theta)=\frac{e^{ikr\cos(\theta-\theta_0)}}{\sqrt{2}}\left(\begin{array}{c} 1\\ e^{i\theta_0} \end{array}\right).
\ee
If $\theta_0=0$, that is, if the particle moves along the x-axis, the wave function will be reduced to $\frac{e^{ikr\cos\theta}}{\sqrt{2}}\left(\begin{array}{c} 1\\ 1 \end{array}\right)$.

\section{General consideration}
To study the scattering induced by  graphene curvature, the Green function was used, which satisfies,
\bee
\left(
\begin{array}{cc}
k   & i\partial_x+\partial_y   \\
i\partial_x-\partial_y   & k
\end{array}
\right) G(\mathbf{r})=
\left(
\begin{array}{cc}
\delta(\mathbf{r})   &  0 \\
0   &  \delta(\mathbf{r})
\end{array}
\right),
\ee
where $k$ is the energy of the carriers with momentum $\mathbf{k}$.

Note that when $r \to \infty$, each component of the Green function should behave as $\frac{e^{ikr}}{\sqrt{r}}$. We finally read the Green function as
\bee
G(r,\theta)=\frac{k}{4}
\left(
\begin{array}{cc}
-iH_0^1(kr)   &  e^{-i\theta}H^1_1(k r) \\
e^{i\theta}H^1_1(k r)    &  -iH_0^1(kr)
\end{array}
\right),
\ee
where $H^1_i(kr)$ is the first kind i'th order Hankel function.

Using the asymptotic behavior of the first kind Hankel functions, we know that
at $ kr \to \infty $,
\bee
G(r,\theta) \to \sqrt{\frac{k}{8\pi }}e^{-\frac{3\pi}{4}i} \frac{e^{ikr}}{\sqrt r}
\left(
\begin{array}{cc}
1   &  e^{-i\theta} \\
e^{i\theta}    &  1
\end{array}
\right).
\ee

Assuming that the incident carriers move along the x-axis, the incident wave function is
$ \psi_i(r^\prime,\theta^\prime)=\frac{e^{ikr^\prime\cos\theta^\prime}}{\sqrt{2}}\left(\begin{array}{c} 1\\ 1 \end{array}\right)$ in position space or
$ \psi_i(\mathbf{p})=\frac{\delta(\mathbf{p}-\mathbf{k}_0)}{\sqrt 2} \left(\begin{array}{c} 1\\ 1 \end{array}\right)$ in momentum space, where $\mathbf{k}_0=(k,0)$ is incident wave vector.

According to the Born approximation, the scattering wave function is
\bee
\psi_s(r,\theta)= \int r^\prime dr^\prime d\theta^\prime G(r,\theta;r^\prime,\theta^\prime) V(\mathbf{r}^\prime) \psi_i(r^\prime,\theta^\prime).
\ee
We further introduce a scattering wave vector $\mathbf{k}_f= \hat{e}_{r}k$. Because we focus only on the limitation $r \to \infty$,
$k|\mathbf{r}-\mathbf{r}^\prime|
\stackrel{r \to \infty}{\to} kr- k\hat{e}_{r}\cdot \mathbf{r}^\prime=kr-\mathbf{k}_f\cdot \mathbf{r}^\prime. $

Regarding the Green function, at the leading order, we have
\bee
G(r,\theta;r^\prime,\theta^\prime) \to \sqrt{\frac{k}{8\pi }}e^{-\frac{3\pi}{4}i} \frac{e^{ikr-i\mathbf{k}_f\cdot\mathbf{r}^\prime}}{\sqrt r}
\left(
\begin{array}{cc}
1   &  e^{-i\theta} \\
e^{i\theta}    &  1
\end{array}
\right).
\ee
By letting $\mathbf{q}=\mathbf{k}_f-\mathbf{k}_0$, we obtain
\beea
\psi_s(r,\theta) &= & \frac{e^{ikr}}{\sqrt{r}} \frac{\sqrt k }{8\sqrt\pi} e^{-\frac{\pi}{4}i}
\left(\begin{array}{cc}
1 & e^{-i\theta} \\
e^{i\theta} & 1
\end{array}\right) \notag \\ &&
\int z_{r\prime}^2dr^\prime d\theta^\prime
\left(
\begin{array}{c}
e^{-i\theta^\prime-i\mathbf{q}\cdot\mathbf{r}^\prime}(\frac{1}{2}+ikr^\prime\cos\theta^\prime)   \\
e^{i\theta^\prime-i\mathbf{q}\cdot\mathbf{r}^\prime}(\frac{1}{2}+ikr^\prime\cos\theta^\prime) \end{array}
\right). \notag \\ &&
\eea
The calculations result in $\psi_s$:
\bee \label{sct1}
\psi_s(r,\theta) =  \frac{e^{ikr}}{\sqrt{r}} \frac{1}{\sqrt 2}
\left(\begin{array}{c} e^{-i\theta/2} \\ e^{i\theta/2}\end{array}\right)F(\theta) \cos\frac{\theta}{2}
,
\ee
where $F(\theta)=\sqrt{\frac{k\pi}{2}}\frac{k}{q} e^{\frac{\pi}{4}i} \int dr^\prime z_{r^\prime}^2 J_1(qr^\prime)$, $J_1(qr)$ is the first-order Bessel function and $q=2k\sin\frac{\theta}{2}$.

Here we emphasize physics interpretations of Eq. (\ref{sct1}).
The factor $\frac{e^{ikr}}{\sqrt{r}}$ stands for circular wave spreading outward, the amplitude of which decreases as $r^{-1/2}$.  The decreasing behavior as $ r^{-1/2}$ but not $r^{-1}$ is attributed to the fact that we are working in two-dimensional space. Furthermore, $\frac{1}{\sqrt 2}\left(\begin{array}{c} e^{-i\theta/2} \\ e^{i\theta/2}\end{array}\right)$ stands for only the particles spreading along $\mathbf{k}_f$, the cut angle between which and the x-axis is $\theta$.

The most important factor is $\cos\frac{\theta}{2}$. Noting that $\cos\frac{\pi}{2}=0$, we conclude that there is no back scattering if a particle passes through a polar-symmetric ripple, at least at the first order. The effective potential $V$ produces a quantum scattering which
changes the pseudo-spin directions of incident particles into other ones. However, the scattering amplitude undergoes a modifying factor, $\cos\frac{\theta}{2}$. Noticing that when $\theta=\frac{\pi}{2}$, $\cos\frac{\theta}{2}=-1$, we think that the factor is due to the pseudospin structure and the chiral conservation of carriers in graphene. According to the Born scattering theory, $F(\theta)\cos\frac{\theta}{2}$ is the differential scattering amplitude along $\theta$-direction.

Now we consider the validity of Eq. (\ref{sct1}). From the perturbation theory, for the Born approximation suitability, we have an enough condition
\bee
||\int d^2 r G(\mathbf{r}_0-\mathbf{r})V(r)|| \ll 1,
\ee
at each $\mathbf{r}_0$, where $||\cdot||$ stands for the matrix norm. Since $H_i^1(kr)=J_i(kr)+iY_i(kr)$ and the incident wave is along x-axis, the above condition reduces into
\bee \label{con1}
\frac{k}{8}|\int d^2 r z_r^2 j(k|\mathbf{r_0-r}|) f(r)|\ll 1,
\ee
where $j(kr)$ may be $J_0(kr)$, $J_1(kr)$, $Y_0(kr)$ or $Y_1(kr)$ and $f(r)$ may be $\frac{1}{2r}$ or $k$.
We assume that the character size and the character height of the ripple is $a$ and $b$ respectively. Therefore, $z_r\sim b/a$.
For the far field, that is, $kr_0\gg 1$ and $r_0>a$, the condition Eq. (\ref{con1}) turns into
\bee
\tag{\ref{con1}$'$}
ka\ll (\frac{8}{\pi})^{2/3}(\frac{a}{b})^{4/3},
\ee
since under normal circumstances, we have $a\gg b$.
For the near field, the most possible divergent point is at $r_0=0$ and it is enough to consider only the inequality
\bee
\tag{\ref{con1}$''$}
\frac{1}{4}\int d r \frac{z_r^2}{r} \ll 1 ,
\ee
where we choose the domain of the integration as $a_0 <r<a$, with the infrared cutoff $a_0 \sim 0.1 nm$, the nearest distance between carbon-carbon atom. For a smoothing ripple, $z_r=0$ at point $r=0$ and $z_r$ drops off rapidly when $r\gg a$, the inequality is always satisfied. However, for a ripple induced by defect, $z_r \neq 0$ at point $r=0$ and the left hand side of the inequality becomes into $\frac{b^2}{4a^2}\ln \frac{a}{a_0}$, which is about $\frac{b^2}{a^2}$ if we choose $a/a_0\sim 10^2$.

In summary, when $ka\ll (\frac{a}{b})^{4/3}$ and $\frac{a}{b}\gg 1$, the Born approximation is valid.
In suspending graphene, $a$ and $b$ are typically at the orders of $10nm$ and  $1 nm$ respectively\cite{sizeheight}. Then, the above condition means that the perturbation theory is valid as long as energies of the scattered carrier are far lower than $3eV$. Noticing that our discussion constraint is also $|E|\ll 3eV$, we conclude that for typical ripples in graphene our perturbation treatment is valid in a quite broad range. However, ripples can be controlled by boundary conditions and making use of graphene¡¯s negative thermal expansion coefficient \cite{nat}. If the character size of the controlled ripple is very large, for instance $a\sim 1\mu m$ and $b\sim 0.1\mu m$. The validity of the theory narrows down to $|E|\ll 10^{-3}eV$, that means that the perturbation theory is nearly invalid in all the energy scale.

In the above discussion we have replaced $(\partial z/\partial r)^2$ by $(b/a)^2$. From the mean value theorem of integrals,  $(\partial z/\partial r)^2$ should be replaced by the mean value, $(\partial z/\partial r)^2|_{r_s}$, where $r_s$ depends on the ripple. We think that, as long as $(\partial z/\partial r)^2$ does not vary sharply, the approximation $(\partial z/\partial r)^2|_{r_s})^2 \sim (b/a)^2$ is correct.

From Eq. (\ref{sct1}), the differential cross section is
\bee \label{cross-section}
\frac{d\sigma(\theta)}{d\theta}=\frac{\pi k^3}{2q^2}\cos^2\frac{\theta}{2}(\int dr z_{r}^2 J_1(qr))^2.
\ee

\section{Examples on scattering}
Next, we study some general properties of the scattering. First, we assume that the amount of energy of the incident particle is very small, $kr \ll 1$. In this situation,
\bee \label{forward}
\frac{d\sigma(\theta)}{d\theta}=\frac{\pi k^3}{8}\cos^2\frac{\theta}{2}(\int dr r z_r^2)^2,
\ee
which means that the differential cross section is nearly isotropic and is proportional to the energy cubed. Therefore, in the quantum scattering process, there is a crucial difference between graphene and ordinary material: in ordinary material, the Born approximation is valid for incident particle with large momentum, whereas in graphene, it is valid for incident particle with small momentum. Note that the above expression is also valid in nearly forward scattering, $\theta \ll \frac{1}{kr}$.

\begin{center}
\begin{minipage}{0.44\textwidth}
\centering
\includegraphics[width=2.2in]{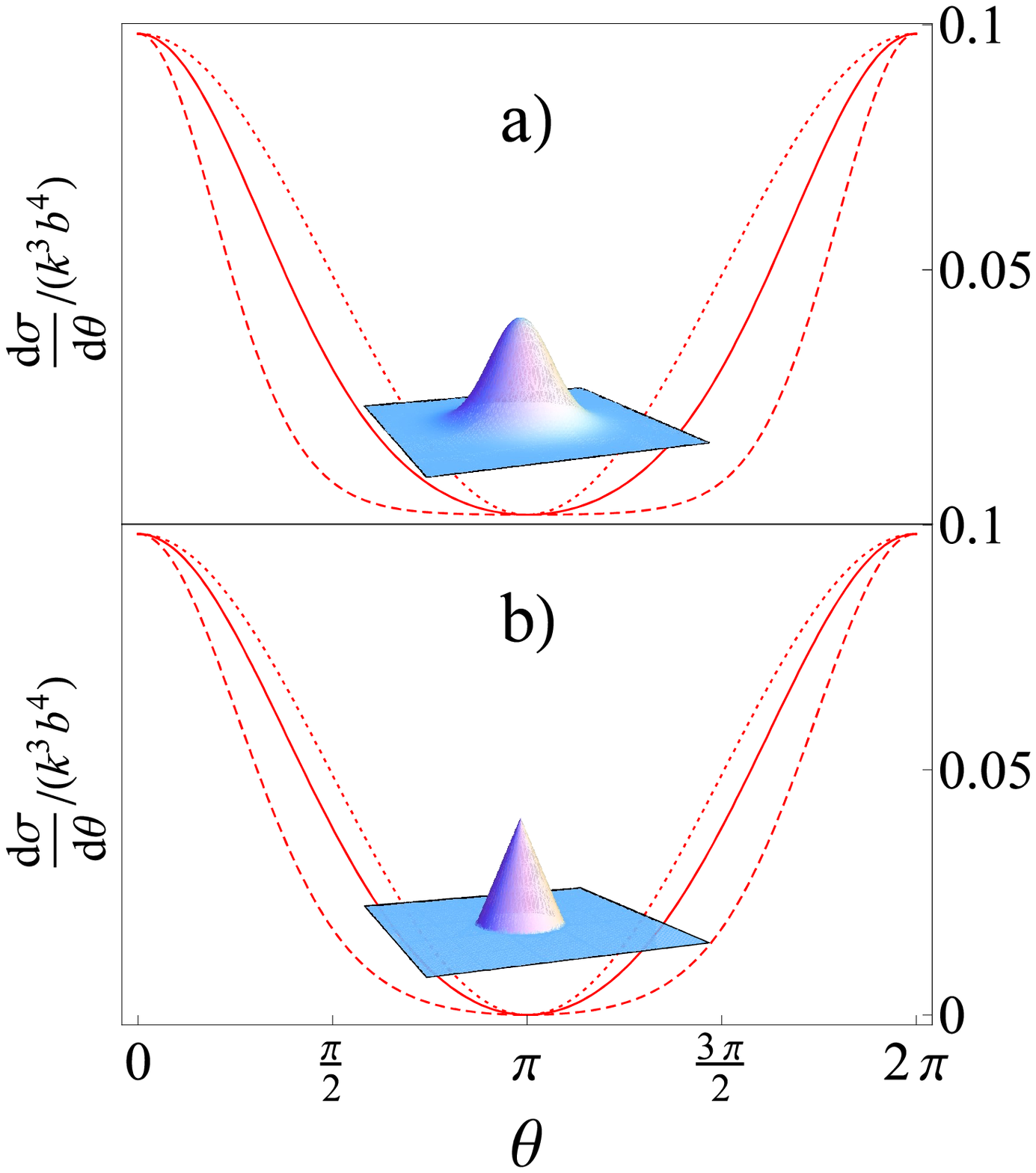}%
\figcaption{Differential cross sections of the Gaussian bump (a) and the defect cone (b) in graphene sheet. Dotted, solid and dashed curves correspond to $ka=0.1$, $ka=1$ and $ka=2$ respectively.
}
\end{minipage}
\setlength{\intextsep}{0.in plus 0in minus 0.1in} 
\end{center}

We now exemplify the quantum scattering process by a Gaussian bump and a cone in graphene sheet. We first assume that the ripple is a Gaussian bump, $z=b e^{-r^2/a^2}$\cite{report,prb07}, where $a$ is the bump radius and $b$ is the bump height.
We have the differential cross section
\bee
\frac{d\sigma(\theta)}{d\theta}=\frac{\pi k^3 b^4}{32}\cos^2\frac{\theta}{2}\exp(-a^2k^2\sin^2\frac{\theta}{2})
\ee
and the total cross section, which is shown in Fig.2,
\bee
\sigma=\frac{\pi^2 k^3 b^4}{32} e^{-a^2k^2/2}(I_0(\frac{a^2k^2}{2})+I_1(\frac{a^2k^2}{2}))
\ee
where $I_i(x)$ is the i'th order of the first kind of modified Bessel function.

For incident particles with small momentum, the differential cross section is isotropic except the pseudospin factor, $\cos^2\frac{\theta}{2}$ and the total cross section is $\sigma_t=\frac{(bk)^3 \pi^2}{32}b$. This  result is interesting, because the cross section does not depend on the size of Gaussian bump, which is in agreement with Eq. (\ref{forward}).

We then consider a most simple case of graphene ripple
$z=
\left\{
\begin{array}{ll}
b(1-r/a) & r<a \\
0 & r>a
\end{array}
\right., $
which is a cone with height $b$ and bottom radius $a$. This situation corresponds to a topological defect in which gauge fields associated with the defect should be applied. Here we focus only on the curvature effect.

We have the differential cross section
$$
\frac{d\sigma(\theta)}{d\theta}=\frac{\pi k^3 b^4}{2q^4 a^4}\cos^2\frac{\theta}{2}(1-J_0(qa))^2,
$$
and a very complex expression for the total cross section, which is also shown in Fig.2.

It is interesting that in a small momentum, the differential cross section is also isotropic, except a pseudospin factor, and the total cross section is $\sigma=\frac{(bk)^3 \pi^2}{32}b$.

\begin{center}
\begin{minipage}{0.44\textwidth}
\centering
\includegraphics[width=2.4in]{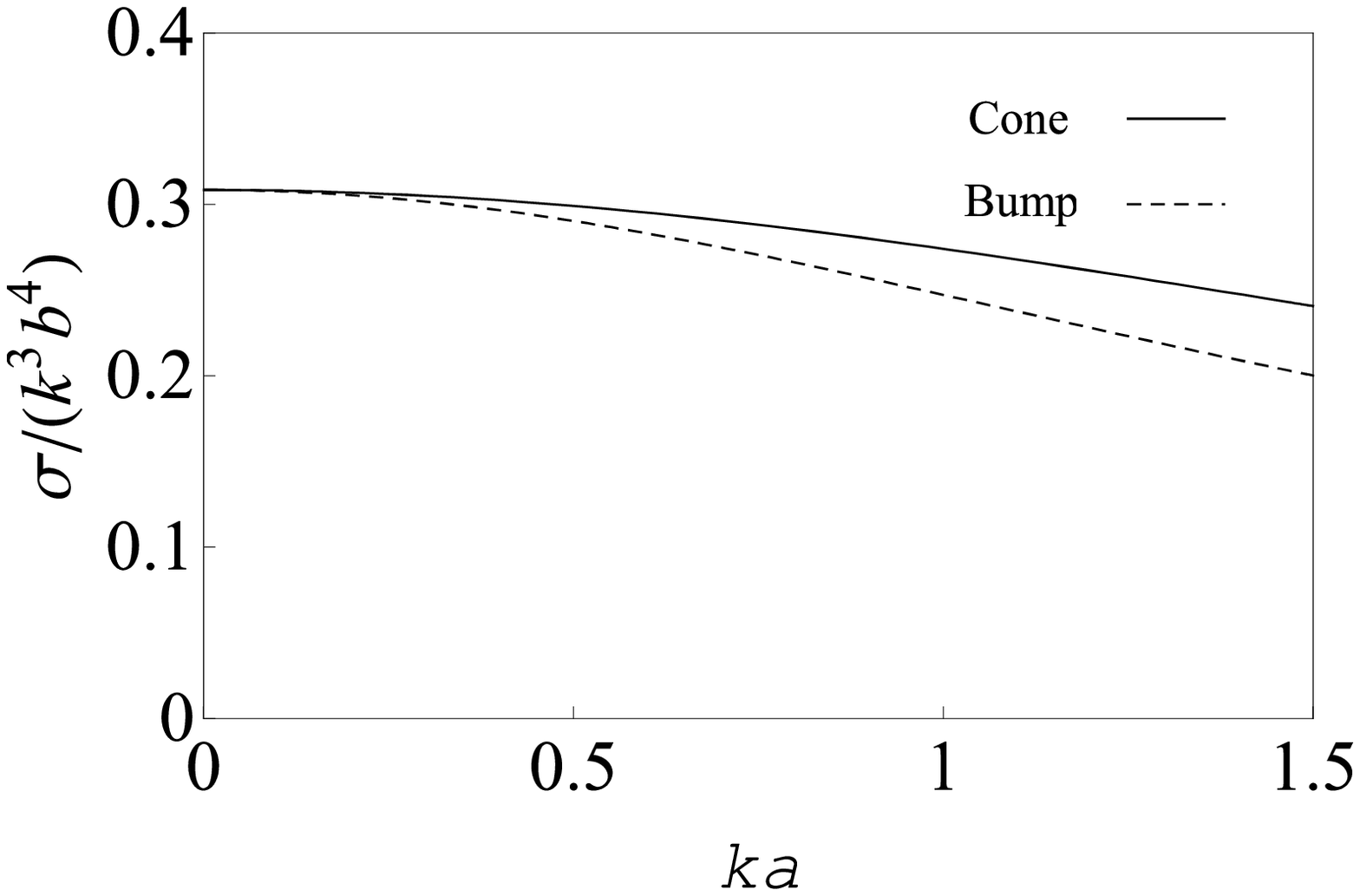}%
\figcaption{ The total cross section for Gaussian bump and cone. The vertical coordinate is the
$\frac{\sigma}{k^3 b^4}$ and the horizontal ordinate is $ka$.
  }
\end{minipage}
\setlength{\intextsep}{0.in plus 0in minus 0.1in} 
\end{center}

Although these two cases differ, they show some similarities. First, the Born approximation is valid in the low-energy scattering process. on the contrary, in ordinary materials,  the Born approximation is generally valid in the high-energy scattering process.
Second, the cross sections are proportional to the energy cubed of the incident particle.
Third, if the energy of the incident particle is very small, the scattering cross section is nearly isotropic, except the pseudospin factor, $\cos\theta/2$, which can be seen in Eq. (\ref{cross-section}).

\section{Summary}
In summary, this paper reports a quantum study on the low-energy scattering process induced by graphene ripples.
We showed that if the graphene is bent only in one direction, its property does not change if a suitable parameter is provided. A general graphene ripple can be equivalent to an effect potential. We also showed a Born approximation of the quantum scattering induced by the (polar symmetric) ripple. However, the result also showed that, unlike in the usual scattering process, the Born approximation used in this study showed an opposite validity but with a quite broad energy range provided the Dirac description of graphene is valid.

Different ripples exhibit different scattering behaviour. However, all the low-energy scattering processes showed some general behaviour. The first is that because of the chiral of the Dirac fermion, the pseudospin factor, $\cos\theta/2$ was shown in scattering amplitude. This factor led to an interesting phenomenon, the suppression of back scattering. Second, the cross section was roughly proportional to the energy cubed of the incident particle. Finally, as long as the energy of the incident particle is small enough, the scattering was nearly isotropic if we ignore the pseudospin factor, $\cos\theta/2$.

The application of graphene may be extended to the design devices based on curved graphene. However, such design needs carefully study of the quantum scattering induced by graphene curvature. It is hoped that the results of this study will be helpful in  designing new type of devices that are based on curved graphene.

A motivation to study ripple graphene and the electron scattering by graphene ripple is that graphene ripple can mimic gravity structure. Here we only consider the elastic properties of graphene. However, there are other types of disorders in graphene, for instance, torsion induced by dislocations. In general relativity, torsion effects are suppressed by the smallness of the gravitational coupling and they are negligible. But, the suppression does not exist in graphene and we can study the torsion effect in experimental and theoretical aspects. In this manuscript we only considered the symmetric part of connection, affine connection. To study torsion effects, one should introduce antisymmetric part of the connection, symbolised by the torsion tensor, which is induced by edge dislocation and proportional to the density of Burgers vector. For the detail study of the torsion effects we refer to Refs. \cite{dislocation,0909,1305}. We think that our treatment is still valid if the antisymmetric part of the connection is induced in Eq. (\ref{affine}). The electron scattering by edge dislocation in 2+1 dimension is an interesting topic and it is worth further research.

\section*{Acknowledgments}
This work is supported by the National Nature Science
foundation of (Grants Nos.51176016 and 11074196).



\balance

\end{multicols}

\end{document}